\documentclass[
 prl,
 preprint,
 amsmath,amssymb,
 aps,
]{revtex4-2}

\usepackage{graphicx}
\usepackage{dcolumn}
\usepackage{bm}

\usepackage{etex}
\usepackage[latin1]{inputenc}
\usepackage{amsmath}
\usepackage[english]{babel}
\usepackage{amsfonts}
\usepackage{amssymb}
\usepackage{esvect}
\usepackage{color,soul}
\usepackage{latexsym}
\usepackage{hyperref}
\newcommand{\ket}{\rangle}

\newcommand{\+}{\dagger}

\newcommand{\kvec}{\mathbf{k}}

\newcommand{\qvec}{\mathbf{q}}

\begin{document}
\title{
Phonon-mediated dark to bright plasmon conversion}

\author{Benjamin Rousseaux}
\email{benjaminrousseaux@gmail.com}
\author{Yanko Todorov}
\author{Angela Vasanelli}
\author{Carlo Sirtori}
\affiliation{Laboratoire de Physique de l'\'Ecole Normale Sup\'erieure, ENS, Universit\'e, PSL, CNRS, Sorbonne Universit\'e, Universit\'e de Paris, F-75005 Paris, France}
\date{\today}
\begin{abstract}
The optical response of a matter excitation embedded in nanophotonic devices is commonly described by the Drude-Lorentz model.  Here, we demonstrate that this widely used approach fails in the case where quantum-confined plasmons of a two-dimensional electron gas  interact strongly with optical phonons. We propose a new quantum model which contains the semiclassical Drude-Lorentz one for simple electronic potentials, but predicts very different results in symmetry-broken potentials. We unveil a new mechanism for the oscillator strength transfer between bright phonon-polariton and dark plasmon modes, enabling thus new quantum degrees of freedom for designing the optical response of nanostructures. 
\end{abstract}
\maketitle
\newpage

\emph{Introduction}--- The emerging field of quantum plasmonics \cite{savage2012revealing,tame2013quantum,zhu2016quantum} explores the strong interaction between quantum emitters and nanoscale plasmonic systems \cite{torma2014strong}. Recently, the ability to design semiconductor heterostructures with a high degree of control over e.g., carrier densities and resonator geometry, has led to semiconductor plasmonics \cite{taliercio2019semiconductor}, which, unlike metals, allows for the design of the effective dielectric response of nanoscale devices. In strongly confined structures, the plasmonic response depends on the quantum properties of single electrons~\cite{vasanelli2020semiconductor}, allowing new designs of infrared emitters and detectors. In the case of polar materials, the physics of these devices is further enriched with the presence of optical phonons, whose interaction with light results in localized and propagating phonon polaritons \cite{caldwell2015low,gubbin2016strong,tamagnone2018ultra,passler2018strong,gubbin2019hybrid,gubbin2021quantum}. Additionally, the well-known Fr\"ohlich interaction between electrons and phonons \cite{frohlich1954electrons} allows for the understanding of the complex interaction between phonons, collective electronic excitations and light, as shown recently  in the framework of a quantum theory of polarons \cite{deliberato2012quantum}, and a perturbative approach for the interaction between intraband electrons and phonon polaritons \cite{franckie2018quantum}.

Another degree of complexity in the physics of nanoscale light-matter interaction is brought by the presence of dark plasmon modes \cite{koh2009electron,liu2009excitation,gomez2013dark,barrow2014mapping,du2019strong,rousseaux2020strong,bitton2020vacuum}, or light-forbidden transitions \cite{cuartero2018light,cuartero2020dipolar}. More specifically, it was shown that an interplay between dark and bright plasmons can be obtained with symmetry-breaking approaches \cite{panaro2014dark,li2018plasmon}. In the context of collective intraband excitations in semiconductor heterostructures, i.e. bulk plasmonic excitations, exploiting dark plasmons could help to design novel detectors in the mid-infrared and terahertz domains.

Here, we present a scheme where dark plasmon modes in a two-dimensional layer are coupled to light via optical phonons in the material. Based on a full quantum model that we derive, we unveil a dark-to-bright plasmon mode conversion mechanism based on the spatial overlap between quantized plasmon microcurrents and phonons. While, in symmetric structures, our predictions are consistent with approaches based on semiclassical plasmon-phonon interaction models, we find that in asymmetric structures plasmon modes that would normally not be visible in optical experiments could be made visible by hydridizing them with phonon polaritons.

This work is organized as follows: we first review the semiclassical interaction between intraband electronic excitations and optical phonons in a semiconductor quantum well, and express the corresponding hybridized plasmon-phonon eigenfrenquencies and oscillator strengths. Next, we introduce a full quantum model based on the diagonalization of the plasmon-phonon part of the light-matter Hamiltonian, and express the corresponding dielectric function in the plasmon-phonon polariton basis. Finally, we study the simple cases of an infinite quantum well, first in a symmetric square well configuration, and second in a symmetry-broken potential well, and show a redistribution of the oscillator strength among all elementary constituents, converting dark plasmon excitations into bright plasmon-phonon resonances. We also explain why the semiclassical approach fails to capture the effect.

\begin{figure}
    \centering
    \includegraphics[width=.4\textwidth]{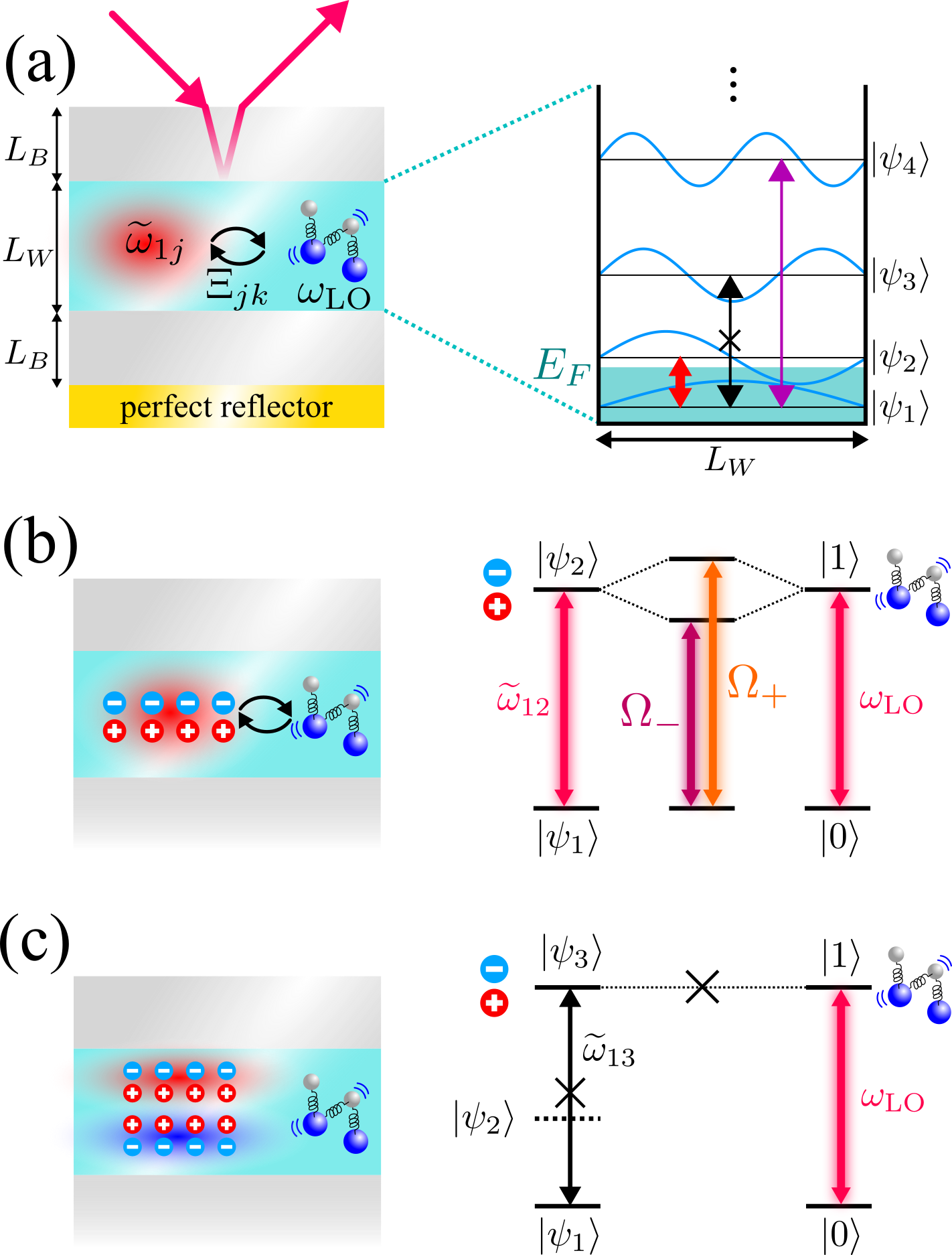}
    \caption{\small (a) System under study: a doped layer of thickness $L_W$ is sandwiched between two barriers of thickness $L_B$, forming a square potential well, on top of a perfect reflector. Wavefunctions (blue lines), and the bright ($\omega_{12}$, $\omega_{14}$) and dark ($\omega_{13}$) transitions are shown as double arrows. (b) Mixing between intersubband plasmons (e.g., transition at frequency $\widetilde{\omega}_{12}$) and phonon polaritons for $\widetilde{\omega}_{12} \approx \omega_\text{LO}$, appearing as plasmon-phonon-polaritons at frequencies $\Omega_\pm$. (c) No coupling is predicted by the semiclassical approach when e.g. the light-forbidden transition $\widetilde{\omega}_{13}$ (crossed double arrow) is resonant with the phonon $\widetilde{\omega}_{13} \approx \omega_\text{LO}$.}
    \label{sketch}
\end{figure}

\emph{Semiclassical plasmon-phonon interaction}---We consider an electron-doped semiconductor layer of thickness $L_W$ with carrier density $N_s$ (in cm$^{-2}$) placed between two undoped layers of thickness $L_B$ having the same high-frequency dielectric constant $\epsilon_\infty$. The top layer is illuminated with a plane wave, of frequency $\omega$ and wavevector $\kvec$, and the structure is assumed to lay on a perfect reflector. The electric field vector of the incident wave is $p$-polarized (or transverse magnetic): it is parallel to the plane of incidence. In this work, $\kvec$ is chosen to be oriented 45 degrees from the $z$-axis.
Due to the reflector, the absorption is then ${\cal A}(\omega) = 1 - {\cal R}(\omega)$, ${\cal R}(\omega)$ being the reflectivity of the whole structure (see Fig. \ref{sketch}(a)). We model the conduction band in the doped layer as an infinite quantum well, whose thickness $L_W$ along the growth axis $z$ is negligibly small compared to the relevant wavelengths, and we assume a constant effective mass $m^*$ (i.e. parabolic dispersion) for the conduction electrons.

For highly doped layers, the classical treatment of the free electron motion is modeled with a Drude dielectric function, giving rise to a Berreman mode in a thin film under oblique incident light \cite{lindau1970experimental,ferrell1958predicted,vassant2012berreman,askenazi2014ultra}. However, when the thickness of the layer becomes comparable to the de Broglie wavelength of electrons, size confinement has to be taken into account by the quantization of electronic energy levels. This results in selection rules for the optically active intraband transitions between subbands $i,j$ within the doped layer.
The solutions of the time-independent Schr\"odinger equation are the single-particle energies $E_i$ and the associated eigenstates $\psi_i(z)$ corresponding to the envelope wavefunctions of conduction electrons in the $z$ direction. In the random-phase approximation \cite{pines2018elementary}, the electronic transitions of frequency $\omega_\alpha = (E_j - E_i)/\hbar,\alpha\equiv i\leftrightarrow j$ are modeled with bosonic excitations $b_{\qvec,\alpha},b^\+_{\qvec,\alpha}$, $\qvec$ being the in-plane wave vector. We restrict our study to a single occupied energy level for simplicity, with the Fermi level $E_F$ laying below the first excited state $|\psi_2\ket$ ($E_F< E_2$, see Fig. \ref{sketch}(a)). Thus, all possible electronic transitions occur between the ground state $|\psi_1\ket$ and excited states $|\psi_j\ket, j>1$ with no transitions occuring between the excited states.
In a perfect square well, the micro-current densities $\xi_{1j}(z) = \psi_1(z)\partial_z\psi_j(z) - \psi_j(z)\partial_z\psi_1(z)$ associated to each transition are even (odd) functions for all transitions $\alpha\equiv 1\leftrightarrow j$ with $j\geqslant 2$ even (odd). Integrating these micro-currents over the growth axis $z$ allows for the determination of the oscillator strengths $f_{1j}$ of the transitions, with all even transitions being bright ($f_{1j}\ne 0$ for $j$ even), while all odd transitions are dark ($f_{1j}= 0$ for $j$ odd). In fact, for an infinite square well, the oscillator strength $f_{12} \simeq 0.96$ of transition $1\leftrightarrow 2$ takes almost all the oscillator strength $\sum_{j=1}^{\infty}{f_{1j}}=1$. The resonance resulting from the collective excitation of electrons in the layer is then usually modeled with a Drude-Lorentz dielectric function in the $z$-direction \cite{liu2000intersubband}: 
\begin{align}
\label{eps12}
\epsilon_{zz}^\text{pl}(\omega) = \epsilon_\infty\left(1 -  f_{12}\frac{\omega_{P12}^2}{\omega^2 - \omega_{12}^2 + i\gamma_\text{pl}\omega}\right),
\end{align}
where $\gamma_\text{pl}$ is the plasmon decay rate.
When higher order transitions must be taken into account, as will be the case in our scheme, Eq. \eqref{eps12} must, however, be revised. A more general derivation of the dielectric function is presented in the Supplementary Information, inspired from previous work for a three- and two-dimensional electron gas \cite{todorov2014dipolar,todorov2015dipolar}.

Transverse optical (TO) phonons in the doped layer are three-dimensional collective modes that oscillate at frequency $\omega_\text{TO}$ for low-valued wavevectors, i.e. around the $\Gamma$ point in the phonon band diagram. They form a dipole-active degenerate continuum of modes that couple to free-space radiation. In the crystal bulk, the associated dielectric function is isotropic:
\begin{align}
\label{epsph}
\epsilon^\text{ph}(\omega) = \epsilon_\infty\left(1 - \frac{R_\text{ph}^2}{\omega^2 - \omega_\text{TO}^2 + i\gamma_\text{ph}\omega}\right),
\end{align}
with $R_\text{ph}^2 = \omega_\text{LO}^2 - \omega_\text{TO}^2$, $\omega_\text{LO}$ being the frequency of longitudinal optical phonons, and $\gamma_\text{ph}$ being the phonon non-radiative rate. In the dielectric functions \eqref{eps12} and \eqref{epsph}, light-matter interaction is accounted for with finite oscillator strengths $f_{12}$ (or implicitly in $R_\text{ph}$ for phonons). The semiclassical dielectric function $\epsilon_{zz}^\text{sc}(\omega)$ is obtained by summing the resonant contributions from plasmons and phonons:

\begin{align}
\label{eps_sc}
\epsilon_{zz}^\text{sc}(\omega) = \epsilon_{zz}^\text{pl}(\omega) + \epsilon^\text{ph}(\omega) - \epsilon_\infty,
\end{align}

This functions allows accounting for an interaction between plasmons and phonons, as its  zeros correspond to the semiclassical plasmon-phonon-polaritons frequencies $\Omega_{\pm,\text{sc}}$ via the biquadratic equation: $(\Omega_{\pm,\text{sc}}^2 - \widetilde{\omega}_{12}^2)(\Omega_{\pm,\text{sc}}^2 - \omega_\text{LO}^2) - f_{12}\omega_{P12}^2R_\text{ph}^2 = 0$.
Such a description of plasmon-phonon interaction basically assumes that the latter is mediated by both plasmon and phonon oscillator strengths, i.e. light-matter couplings (see Fig. \ref{sketch}(b)). Thus, oscillators with very weak light-matter couplings have a negligible contribution in the overall semiclassical dielectric function, as illustrated in Fig. \ref{sketch}(c). Inverting the $z$-component of the semiclassical dielectric tensor, we obtain a relation of the form:
\begin{align}
\label{epssc}
\frac{\epsilon_\infty}{\epsilon_{zz}^\text{sc}(\omega)} = 1 + \sum_{\lambda = \pm}\frac{{\cal R}_{\lambda,\text{sc}}^2}{\omega^2 - \Omega_{\lambda,\text{sc}}^2 + i\gamma_{\lambda,\text{sc}}\omega},
\end{align}
with the plasmon-phonon polaritonic effective plasma frequencies ${\cal R}_{\lambda,\text{sc}}$ (including their oscillators strengths) and decay rates $\gamma_{\lambda,\text{sc}}$. The inversion of the dielectric function, as in expression \eqref{epssc}, corresponds to a new function whose poles are the polaritonic frequencies $\Omega_{\lambda,\text{sc}}$ in the limit of vanishing dissipation rates $\gamma_{\lambda,\text{sc}}$. This expression, being the usual coupled Lorentz oscillators dielectric function, will be useful to be compared with the new dielectric function we derive in the next section.

\emph{Full quantum model}---In the previous section, we derived the semiclassical plasmon-phonon interaction by first expressing the plasmon dielectric function $\epsilon^\text{pl}(\omega)$ (Eq. \eqref{eps12}), then the optical phonon dielectric function $\epsilon^\text{ph}(\omega)$ (Eq. \eqref{epsph}), and by mixing both of them in Eqs. \eqref{eps_sc}--\eqref{epssc}. Due to the relation between light-matter Hamiltonians and corresponding dielectric functions \cite{todorov2014dipolar,todorov2015dipolar}, it seems natural to wonder whether treating the full plasmon-phonon Hamiltonian and express its global dielectric function should yield the same result as Eqs. \eqref{eps_sc} or \eqref{epssc}. We now express the full light-matter Hamiltonian $H=H_\text{mat} + H_\text{light} + H_\text{light-mat}$ in terms of plasmon-phonon polaritons, which are mixed matter excitations, and establish the related dielectric function. The full matter Hamiltonian $H_\text{mat}$ is derived in the Supplementary Information and has the form:
\begin{align}
\label{Hmat}
{H}_\text{mat} &= \sum_{\qvec,j}\hbar\widetilde{\omega}_{1j}p_{\qvec j}^\+p_{\qvec j} + \sum_{\qvec}\hbar\omega_\text{LO}\Big(\sum_kr_{\qvec k}^\+r_{\qvec k}+ s_{\qvec}^\+s_{\qvec}\Big) \nonumber\\ &+ \sum_{\qvec,j,k}\frac{\hbar\Xi_{jk}}{2}\Big(p_{\qvec j}^\+ + p_{-\qvec j}\Big)\Big(r_{-\qvec k}^\+ + r_{\qvec k}\Big),
\end{align}
where $\widetilde{\omega}_{1j} = (\omega_{1j}^2+\omega_{P1j}^2)^{1/2}$, $p_{\qvec j},p_{\qvec j}^\+$ are the annihilation and creation operators of the intersubband plasmon associated with the transition $1\leftrightarrow j$, $r_{\qvec k},r_{\qvec k}^\+$ are operators associated to plasmon-coupled phonons and  $\Xi_{jk}$ are the plasmon-phonon coupling strengths. Since the phonon excitations form a continuum of plane waves, the phonon polarization vector is spanned over all possible spatial harmonics. However, only the phonon spatial harmonics that are matched to the shape of the electronic quantum-confined  microcurrents couple to the plasmon excitations. The phonon modes not coupled to the plasmons are accounted through the bosonic operators  $s_\qvec,s_\qvec^\+$. Next, to diagonalize the Hamiltonian, we introduce the polaritonic operators $\Pi_{\qvec \lambda} = \sum_{j}(x_{\lambda j}p_{\qvec j} + y_{\lambda j}p_{-\qvec j}^\+) + \sum_{k}(m_{\lambda k} r_{\qvec k} + h_{\lambda k}r_{-\qvec k}^\+)$, with the Hopfield coefficients $x_{\lambda j},y_{\lambda j},m_{\lambda k},h_{\lambda k}$ and new indices $\lambda$ labeling the polaritonic modes. These operators satisfy the eigenvalue problem $\big[\Pi_{\qvec\lambda},{H}_\text{mat}\big] = \Omega_\lambda\Pi_{\qvec\lambda}$, where $\Omega_\lambda$ are the eigenfrequencies of the polariton modes, and the Hamiltonian is expressed in the new basis:
${\cal H}_\text{mat} = \sum_{\qvec, \lambda} \hbar\Omega_{\lambda}\Pi_{\qvec \lambda}^\+\Pi_{\qvec \lambda} + \sum_\qvec\hbar\omega_\text{LO}s_\qvec^\+s_\qvec$. The final step consists in finding the light-matter coupling strengths ${\cal R}_\lambda$ of the plasmon-phonon polaritons, which we identify in an expansion of the light-matter interaction term ${\cal H}_\text{light-mat}$ in the polaritonic basis. We find that the $z$-component of the dielectric tensor accounting for size confinement is:
\begin{align}
\label{queps}
\frac{\epsilon_\infty}{\epsilon_{zz}^\text{qu}(\omega)} = 1 &+ \sum_\lambda\frac{{\cal R}_\lambda^2}{\omega^2 - \Omega_\lambda^2 + i\gamma_\lambda\omega} \nonumber\\
&+ \Big(1 - \sum_k\eta_k\Big)\frac{R_\text{ph}^2}{\omega^2 - \omega_\text{LO}^2 + i\gamma_\text{ph}\omega},
\end{align}
where the polaritonic decay rates $\gamma_\lambda$ are determined from the Hopfield coefficients and we introduced the projected phonon `oscillator strengths' $\eta_k$, corresponding to interactions with plasmons. It is clear that the dielectric function in Eq. \eqref{queps} generally differs from \eqref{epssc}: the light-matter couplings ${\cal R}_\lambda$ correspond to linear combinations mixing the original plasmon and phonon oscillator strengths $R_{1j}=f_{1j}\omega_{P1j}^2$ and $R_\text{ph}$, allowing dark transitions to hybridize with bright phonons. This differs dramatically from the semiclassical couplings ${\cal R}_{\lambda,\text{sc}}$ that vanish if either of the plasmon or phonon oscillator strength is zero. To further underline this difference, we derive, in the Supplementary Information, the effective matter Hamiltonian corresponding to the semiclassical dielectric function, Eq. \eqref{eps_sc}, for the case of transition $1\leftrightarrow 2$. We show that the plasmon-phonon coupling strength differs from that in \eqref{Hmat} by a factor $\propto\sqrt{f_{12}}\approx 1$, which explains both why the semiclassical approach captures well the physics for bright transitions, but fails to describe coupling with dark ones. 

In addition, the last term in Eq. \eqref{queps} corresponds to remaining phononic excitations that do not couple to plasmons, since the factor $\eta_k$ quantifies the proportion of phonons with wavenumber $k$ interacting with plasmons. The factor $1-\sum_k\eta_k$ factor thus accounts for the  spatial mismatch between three-dimensional phonons and the $xy$-plane confined plasmons, as explained above. This factor is absent from the semiclassical theory, which assumes a perfect spatial overlap between plasmons and phonons; the latter can be achieved only for phonons coupled with bulk plasmon excitations. The detailed derivation of the quantized plasmon-phonon interaction as well as the dielectric tensor are provided in the Supplementary Information. In the following, we compare absorption spectra resulting from the well-known semiclassical expression \eqref{epssc} and our quantum approach \eqref{queps}. We specifically focus, in this paper, on the ability to couple dark plasmon modes to light by hydridizing them with bright phonons, but a more detailed study based upon our newly derived model will be presented in future work.
\begin{figure}
    \centering
\includegraphics[width=.45\textwidth]{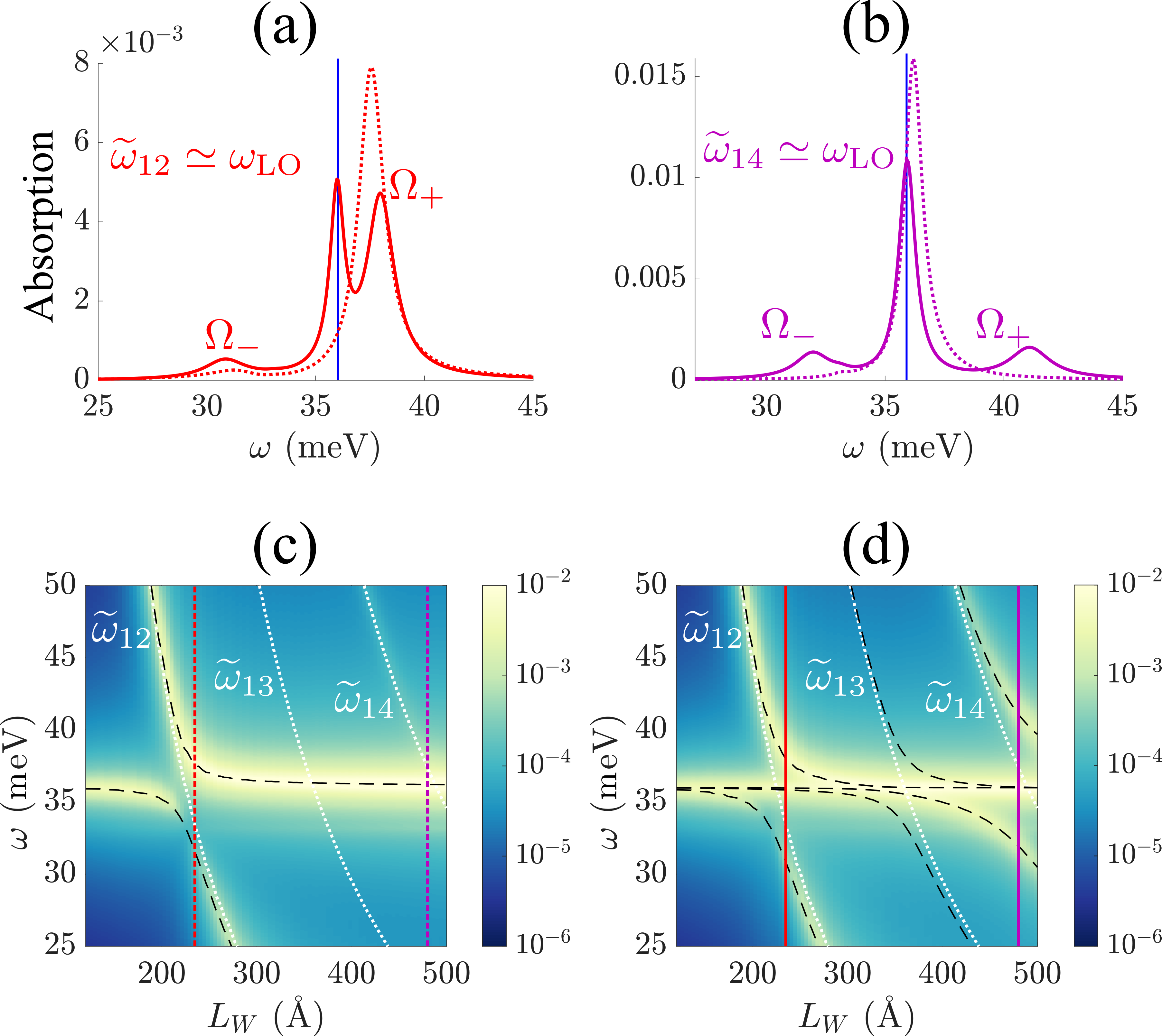}
    \caption{\small Plasmon-phonon absorption spectra for an infinite GaAs square well of thickness $L_W$. (a) Absorption spectra for semiclassical (dotted) and quantum (solid) models, the bright plasmon $\widetilde{\omega}_{12}$ being resonant with $\omega_\text{LO}$ (vertical line), corresponding to $L_W = 235$ \AA. (b) Same as (a) for $L_W = 480$ \AA, matching the plasmon $\widetilde{\omega}_{14}$ with $\omega_\text{LO}$. (c) Semiclassical (log scale) and (d) full quantum absorption versus $\omega$ and $L_W$. (a) and (b) spectra (vertical lines), bare plasmon energies (white dotted lines), $\Omega_{\lambda,\text{sc}}$ the semiclassical model and $\Omega_\lambda$ energies (black dashed lines) are shown in (c) and (d), respectively.}
    \label{1214}
\end{figure}
\begin{figure*}
    \centering
    \includegraphics[width=.9\textwidth]{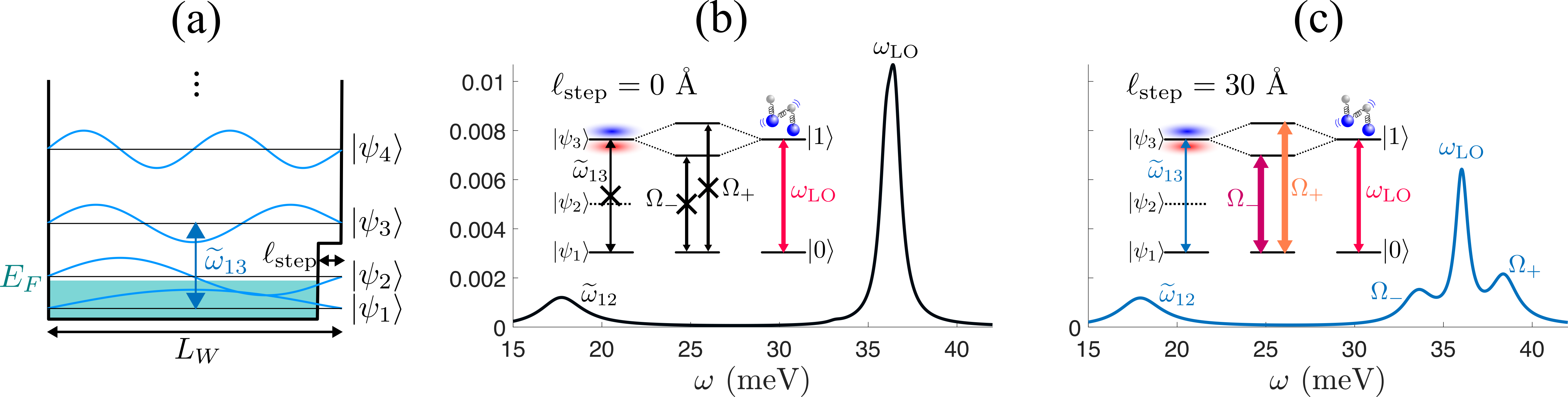}
    \caption{Dark to bright plasmon conversion in GaAs for $N_s = 2.4\times 10^{11}$ cm$^{-2}$. (a) Potential well of the doped layer of total length $L_W$ and with a small step of length $\ell_\text{step}$ and height $h_\text{step}$. (b) Absorption spectrum computed from the quantum approach, for $\widetilde{\omega}_{13} \simeq\omega_\text{LO}$ ($L_W = 368$ \AA) and no step ($\ell_\text{step} = 0$ \AA). (c) Absorption spectrum for $\ell_\text{step}=30$ \AA\ and $h_\text{step} = 30$ meV. The insets in (b) and (c) illustrate the corresponding level hybridization sketches.}
    \label{13}
\end{figure*}

\emph{Results}---We study numerically the absorption spectra of an infinite quantum well consisting of a doped gallium arsenide (GaAs) layer of thickness $L_W$ sandwiched between two barriers of thickness $L_B$ and permittivity $\epsilon_\infty$. To obtain the parameters of the dielectric function \eqref{queps}, we solve Schr\"odinger's equation for the envelope functions in the quantum well, and numerically perform the multiple Hopfield-Bogolyubov diagonalizations described in the previous section and in the Supplementary Information. The electronic density per unit area $N_s = 1.5\times 10^{11}$ cm$^{-2}$ is kept unchanged for all of our calculations, restricting the well thickness $L_W$ to values below 500 \AA\ and keeping the Fermi level below the first excited state $E_F<E_2$, so we do not enter the regime where multiple levels are populated. We can then restrict our study to transitions $1\leftrightarrow j$, with $j = 2,3,4$, even though we fully account for 50 levels in the well to ensure convergence. 

In Fig. \ref{1214}(a), we plot the absorption spectrum of the structure by matching the frequency of the bright intersubband plasmon to that of the LO phonon: $\widetilde{\omega}_{12}\simeq\omega_\text{LO}$, and for this we set $L_W = 235$ \AA. Firstly, we notice that our quantum model (solid line) predicts plasmon-phonon polariton peaks whose position are very close to the semiclassical ones (dotted line): $\Omega_{\lambda}\simeq\Omega_{\lambda,\text{sc}}$. However, significant differences between the semiclassical and quantum oscillator strengths (peak heights), ${\cal R}_\lambda \ne {\cal R}_{\lambda,\text{sc}}$, are found. Another feature that is clearly absent in the semiclassical prediction is the presence of a residual LO phonon peak at $\omega = \omega_\text{LO}$, corresponding to phonons that are spatially mismatched with the bright plasmon, and hence do not couple to it (last term in Eq. \eqref{queps}). 

In Fig. \ref{1214}(b), we match the low-radiative plasmon corresponding to transition $1\leftrightarrow 4$ to the LO phonon, by setting the well length $L_W = 480$ \AA. This time, the semiclassical approach only shows the bare LO phonon peak, because the oscillator strength of the plasmon is so low that the model predicts negligible coupling with the phonon. The situation is radically different from the prediction of the quantum model, with the appearance of two polaritonic peaks at $\Omega_\lambda,\lambda=\pm$. This is explained by the coupling strength $\Xi_{jk}$ between the plasmon and the phonon being a function of the plasma frequency $\omega_{P14}$ \emph{and not} a function of its oscillator strength, the latter quantifying its coupling with light. Therefore, because of hybridization between the LO phonon, which is bright, and the hardly radiative $1\leftrightarrow 4$ plasmon, the oscillator strength of the phonon is redistributed among the polaritonic modes and the residual (uncoupled) phonon.

Figs. \ref{1214}(c) and (d) present maps of the absorption spectra in log scale for the semiclassical and quantum approaches, respectively. Avoided crossing behaviours are found in both maps, as expected, between the bright $1\leftrightarrow 2$ plasmon and phonons, but only the quantum approach displays an avoided crossing  between the low-radiative $1\leftrightarrow 4$ plasmon and phonons. In (d), we find a Rabi splitting $\sim 9$ meV, suggesting that plasmon-phonon polaritons are on the onset of ultrastrong coupling regime with $\Xi_{14}/\omega_\text{LO} \sim 0.12$. Noticeably, an anti-crossing behaviour is also found in the behaviour of the eigenfrequencies $\Omega_\lambda$ between the dark $1\leftrightarrow 3$ plasmon and phonons, but with no oscillator strength, hence no visible peak in the absorption spectrum. This feature is explained by the fact that phonons will hybridize only with plasmons whose spatial distribution (i.e. the micro-current distribution $\xi_{1j}(z)$) have significant spatial overlap with the phonons themselves. In the case of an infinite well, the micro-current densities $\xi_{1j}(z)$ are always even (odd) for $j$ even (odd). Therefore, dark plasmons having odd distributions, they hydridize only with phonons whose spatial distribution is also odd, hence the hydridization remains light-forbidden (see Supplementary Information). Analogously, light-allowed plasmons hydridize only with bright phonons with an even spatial distribution. This coupling mechanism depending on the parity of the uncoupled microcurrents is reminiscent of Fermi resonances seen in the infrared spectra of vibrating molecules, involving the coupling between molecular vibrational modes having the same symmetries \cite{albert2011fundamentals}.

The symmetry of the square quantum well implies that wavefunctions $\psi_j(z)$, and by extension plasmonic micro-currents $\xi_{1j}(z)$, are either even or odd. Since odd micro-currents generate dark plasmons that are not exploitable in optical experiments, it is beneficial to break the symmetry of the well so that the distinction between even and odd functions disappears. Usually, symmetry-breaking has to be significant in order to convert dark modes to bright ones (see e.g. Ref. \cite{panaro2014dark}). In the following, we show that a tiny asymmetry is sufficient to exploit dark modes and fully convert them, thanks to the plasmon-phonon interaction. In the spirit of our former investigation, we consider a similar device, with the following changes: the electronic density is increased: $N_s = 2.4\times 10^{11}$ cm$^{-2}$ and the GaAs potential well of total thickness $L_W$ is this time asymmetric and present a small step of length $\ell_\text{step} \ll L_W$. Since the light-forbidden transition $1\leftrightarrow 3$ has no net dipole moment, we set the step height $h_\text{step}$  between $E_2$ and $E_3$ (see Fig. \ref{13}(a)) and we parametrize the well so that $\widetilde{\omega}_{13}\simeq\omega_\text{LO}$. Fig. \ref{13}(b) shows the absorption spectrum for $L_W = 368$ \AA\ and no step (square well), while Fig.\ref{13}(c) shows the absorption spectrum obtained for $\ell_\text{step}=30$ \AA\ and $h_\text{step}=30$ meV. Our quantum model shows that the (formerly) dark plasmon hydridizes with bright phonons,in contrast with the symmetric case shown in \ref{13}(b), where only a single peak at $\omega_\text{LO}$ and the detuned bright plasmon at $\widetilde{\omega}_{12}$ are revealed. Unlike the results shown for the symmetric well, the dark plasmon-phonon hydridized modes are now light-allowed, despite the tiny oscillator strength ratio $f_{13}/f_{12} = 2.68\times 10^{-8}$ and due to symmetry-breaking induced by the small step. We further study the role of symmetry-breaking in the Supplementary Information, where the step lengths and heights are varied, showing that the splitting is robust even for small defects. The ratio $f_{13}/f_{12}$ is shown to not exceed 0.1 for $\ell_\text{step} = L_W/2$, and is orders of magnitude smaller for $\ell_\text{step}\ll L_W$, confirming that this effect is indeed a redistribution of oscillator strength between the dark plasmon and the phonon-polaritons.

\emph{Conclusion}---In summary, we have presented a new model taking into account the effects of size confinement of electrons in plasmon-phonon interaction in semiconductor materials. This model explores plasmon-phonon coupling schemes beyond the semiclassical Drude-Lorentz coupled oscillators model. Phonon modes are automatically sorted between those overlapping spatially with the plasmon modes, and the remaining ones being uncoupled to plasmons. But the most interesting toolbox is the ability to describe plasmon-phonon couplings via direct dipole-dipole interaction, quantified by the effective plasma frequencies, rather than assuming a coupling strength proportional to a product of their oscillator strengths, hence assuming that plasmon-phonon interaction is mediated only by radiation. Rather, hydridization occurs between them and oscillator strengths are redistributed, opening the path to dark-to-bright plasmon conversion, as we have demonstrated. We hope this model will help in the understanding and design of novel devices in nanophotonics, mid-infrared sources and detectors, and resonance engineering with nanodevices.

\begin{acknowledgments}
The authors acknowledge support from the Chaire ENS/CNRS-Thales, the ERC project UNIQUE and Agence Nationale de la Recherche (Grant No. ANR-19-CE30-0032-01).
\end{acknowledgments}

\bibliography{main.bbl}

\end{document}